\documentstyle[12pt]{article}
\begin{document}
\centerline{\Large\bf Palatini approach to Born-Infeld-Einstein theory}
\centerline{\Large\bf and a geometric description of electrodynamics}
\vskip .7in
\centerline{Dan N. Vollick}
\centerline{Department of Physics and Astronomy}
\centerline{and}
\centerline{Department of Mathematics and Statistics}
\centerline{Okanagan University College}
\centerline{3333 College Way}
\centerline{Kelowna, B.C.}
\centerline{V1V 1V7}
\vskip .9in
\centerline{\bf\large Abstract}
\vskip 0.5in
The field equations associated with the Born-Infeld-Einstein action are 
derived using the Palatini variational technique. In this approach 
the metric and connection are varied independently and 
the Ricci tensor is generally 
not symmetric. For sufficiently small curvatures the resulting
field equations can be divided into two sets.
One set, involving the antisymmetric part of the Ricci tensor
$R_{\stackrel{\mu\nu}{\vee}}$,
consists of the field equation for a massive vector field. The other set
consists of the Einstein field equations with an energy
momentum tensor for the vector field plus additional corrections.
In a vacuum with $R_{\stackrel{\mu\nu}{\vee}}=0$ the field equations
are shown to be the usual Einstein vacuum equations. This extends
the universality of the vacuum Einstein equations, discussed by
Ferraris et al. \cite{Fe1,Fe2}, to the Born-Infeld-Einstein action.
In the simplest version of the theory there is a single coupling constant
and by requiring that the Einstein field equations hold to a good
approximation in neutron stars
it is shown that mass of the vector field exceeds the lower bound 
on the mass of the photon.
Thus, in this case the vector field cannot represent the 
electromagnetic field and would describe a new geometrical field.
In a more general version in which the symmetric and antisymmetric
parts of the Ricci tensor have different coupling constants it is possible
to satisfy all of the observational constraints if the antisymmetric coupling
is much larger than the symmetric coupling.
In this case the 
antisymmetric part of the Ricci tensor can describe the electromagnetic
field.
\newpage
\section*{Introduction}
In the 1930's Born and Infeld \cite{Bo1} attempted to eliminate 
the divergent self energy of the electron by modifying Maxwell's
theory.
Born-Infeld electrodynamics follows from the Lagrangian 
\begin{equation}
L=-\frac{1}{4\pi b}\left\{\sqrt{-det(g_{\mu\nu}+bF_{\mu\nu})}
-\sqrt{-det(g_{\mu\nu})}\right\}\; ,
\end{equation}
where $g_{\mu\nu}$ is the metric tensor and $F_{\mu\nu}$ is the
electromagnetic field tensor.
In the weak field limit this Lagrangian reduces to the
Maxwell Lagrangian plus small corrections. For strong fields the
field equations deviate significantly from Maxwell's theory and 
the self energy of the electron can be shown to be finite.
The Born-Infeld action also appears in string theory. The action
for a D-brane is of the Born-Infeld form with two fields, a
gauge field on the brane and the projection of the Neveu-Schwarz
B-field onto the brane \cite{Po1}.
   
In the same spirit one
can attempt to modify the Lagrangian of general relativity from
$L=-(1/2\kappa) R$, where $R$ is the Ricci scalar, to
\begin{equation}
L=-\frac{1}{\kappa b}\left\{\sqrt{-det(g_{\mu\nu}+bR_{\mu\nu})}
-\sqrt{-det(g_{\mu\nu})}\right\}+L_M
\label{intro}
\end{equation}
where $R_{\mu\nu}$ is the Ricci tensor, $L_M$ is the matter Lagrangian 
and $\kappa=8\pi G$.
Deser and Gibbons \cite{De1} considered this type of Lagrangian 
and derived the equations of 
motion by using a purely metric variation, i.e. 
they took the connection to be the Christoffel symbol. This leads
to fourth order equations with ghosts. To eliminte the ghosts they
considered modifying the action so that the quadratic terms in 
the Taylor series expansion of the Lagrangian vanished. 
Feigenbaum, Freund and Pigli \cite{Fe1} and Feigenbaum \cite{Fe2} have 
also examined Born-Infeld like gravitational actions in two and
four dimensions using a purely metric variation. They found that the
form of the action produces a curvature limiting effect and that the
spacetime inside a black hole is nonsingular.
   
Here I take a different approach and derive the field equations
using a Palatini variation. In the Palatini approach the connection
and metric are treated as independent variables. If the action is
taken to be $L=-(1/2\kappa) R+L_M$ the variation with respect to the
connection fixes the connection to be the Christoffel symbol.
The variation with respect to the metric then gives the
Einstein field equations. However, the purely metric variation and
the Palatini variation give different field equations for Lagrangians
that are general functions
of the Riemann tensor and its contractions. For example if $L=f(R)+
L_M$ it can be shown \cite{Vo1} that the field equations derived from
the Palatini approach are second order in contrast to the fourth order
equation that follow from a purely metric variation.  
If the curvature is is much less than $1/b$ the field equations that 
follow from 
(\ref{intro}) describe a massive vector field, with $R_{\stackrel{\mu\nu}
{\vee}}$ being proportional to the electromagnetic field tensor, 
and the Einstein 
equations plus small corrections. The energy momentum tensor for
the massive vector field automatically appears in the Einstein
equations. 
  
In a vacuum with $R_{\stackrel{\mu\nu}{\vee}}=0$ the equations reduce to
the usual vacuum Einstein field equations. This extends the universality
of the vacuum Einstein field equations discovered by Ferraris et al.
\cite{Fe1,Fe2} to the Born-Infeld-Einstein action. This implies that there
is no curvature limiting effect and that black holes will contain a 
spacetime singularity. There are however, corrections to Einstein's 
equations in interior regions. If these corrections are taken to 
be small in neutron stars the mass of the vector field
exceeds the lower limit on the mass of the photon. Thus, if this theory
is realized in nature it describes a new massive field.
  
On a manifold with a general connection there are two nonzero contractions
of the Riemann tensor. One can consider including both contractions in the
Born-Infeld-Einstein Lagrangian and this theory will then have two 
coupling constants. One 
coupling is associated with the symmetric part of the Riemann tensor and
the other is associated with the antisymmetric part. If the antisymmetric
coupling is much larger than the symmetric coupling all the
constraints can be satisfied and the antisymmetric part of the Ricci 
tensor can describe the electromagnetic field. 
\section*{The Field Equations I} 
The field equations for the theory follow from the Born-Infeld-Einstein
action
\begin{equation}
L=-\frac{1}{\kappa b}\left\{\sqrt{-det(g_{\mu\nu}+bR_{\mu\nu})}
-\sqrt{-det(g_{\mu\nu})}\right\}+L_M
\label{lag}
\end{equation}
where $R_{\mu\nu}$ is the Ricci tensor, $\kappa=8\pi G$ 
and $L_M$ is the matter Lagrangian. The Ricci tensor is given by
\begin{equation}
R_{\mu\nu}=\partial_{\nu}\Gamma^{\alpha}_{\mu\alpha}-\partial_{\alpha}
\Gamma^{\alpha}_{\mu\nu}-\Gamma^{\alpha}_{\beta\alpha}\Gamma^{\beta}_{\mu\nu}
+\Gamma^{\alpha}_{\beta\mu}\Gamma^{\beta}_{\alpha\nu}
\label{Ricci1}
\end{equation}
and the connection is taken to be symmetric. Note that $R_{\mu\nu}$ is not
symmetric in general. If the curvature is much smaller than $b^{-1}$ an 
expansion of (\ref{lag}) gives the Einstein Lagrangian to lowest order in
$b$. Thus, for sufficiently weak fields Einstein's equations will hold to
a good approximation.
   
Varying the action with respect to $g_{\mu\nu}$ gives
\begin{equation}
\sqrt{P}\left(P^{-1}\right)^{\underline{\mu\nu}}-\sqrt{g}g^{\mu\nu}
=\sqrt{g}\kappa bT^{\mu\nu}
\label{eq1}
\end{equation}
where $P_{\mu\nu}=g_{\mu\nu}+bR_{\mu\nu}$,  $P^{-1}$ is the inverse
of $P$, $(P^{-1})^{\underline{\mu\nu}}$
is the symmetric part of $P^{-1}$, $P=-det(P_{\mu\nu})$ and $g=-det(g_{\mu\nu})$.
Varying with respect to $\Gamma^{\alpha}_{\mu\nu}$ gives
\begin{equation}
\nabla_{\alpha}\left[\sqrt{P}\left(P^{-1}\right)^{\underline{\mu\nu}}\right]
-\frac{1}{2}\nabla_{\beta}\left\{\sqrt{P}\left[\delta^{\mu}_{\alpha}
\left(P^{-1}\right)^{\beta\nu}+\delta^{\nu}_{\alpha}\left(P^{-1}
\right)^{\beta\mu}\right]\right\}=0
\label{eq2}
\end{equation}
and contracting over $\alpha$ and $\nu$ gives
\begin{equation}
\nabla_{\alpha}\left[\sqrt{P}\left(P^{-1}\right)^{\stackrel{\alpha\mu}
{\vee}}\right]=-\frac{3}{5}\nabla_{\alpha}\left[\sqrt{P}\left(P^{-1}
\right)^{\underline{\alpha\mu}}\right]\;
\label{anti}
\end{equation}
where $(P^{-1})^{\stackrel{\mu\nu}{\vee}}$ is the antisymmetric part of
$P^{-1}$.
  
It will be instructive to consider the vacuum equations when 
$R_{\mu\nu}$ is symmetric. Since the antisymmetric part of $R_{\mu\nu}$
is to be proportional to the vector field this will describe
a spacetime free of all matter, including the vector field. In this
case the left hand side of (\ref{anti}) vanishes implying that
\begin{equation}
\nabla_{\alpha}\left[\sqrt{P}\left(P^{-1}\right)^{\underline{\mu\nu}}
\right]=0.
\label{eq2a}
\end{equation}
Substituting equation (\ref{eq1})
into equation (\ref{eq2a}) gives
\begin{equation}
\nabla_{\alpha}\left(\sqrt{g}g^{\mu\nu}\right)=0
\end{equation}
and this tells us that the connection is the Christoffel symbol. Taking
the determinant of (\ref{eq1}) gives $p=g$ and we see that 
\begin{equation}
R_{\mu\nu}=0.
\label{Ricci}
\end{equation}
Thus, in a vacuum the resulting field equations are the Einstein field
equations, independent of the value of b.
This extends the universality of the Einstein vacuum equations
discovered by Ferraris et al. \cite{Fe1,Fe2} to Born-Infeld like actions.
In their papers Ferraris et al. showed that Lagrangains of the form
$L=F(R)$ and $L=f(R^{\mu\nu}R_{\mu\nu})$ always give the Einstein
vacuum equations under a Palatini variation.

Now go back to the general case where $R_{\mu\nu}$ is not 
symmetric. From equation (\ref{anti}) we see that
\begin{equation}
\nabla_{\alpha}\left[\sqrt{P}\left(P^{-1}\right)^{\alpha\mu}\right]=
\frac{2}{5}\nabla_{\alpha}\left[\sqrt{P}\left(P^{-1}\right)^
{\underline{\alpha\mu}}\right]\; .
\label{eq2b}
\end{equation}
Substituting equations (\ref{eq1}) and (\ref{eq2b}) into (\ref{eq2}) gives
\begin{equation}
\nabla_{\alpha}\left[\sqrt{g}\left(g^{\mu\nu}+\kappa bT^{\mu\nu}\right)
\right]-\frac{1}{5}\nabla_{\beta}\left\{\sqrt{g}\left[\left(g^{\mu\beta}+\kappa
bT^{\mu\beta}\right)\delta^{\nu}_{\alpha}+\left(g^{\nu\beta}+\kappa b
T^{\nu\beta}\right)\delta^{\mu}_{\alpha}\right]\right\}=0.
\label{last}
\end{equation}
Since the trace of the above system of equations vanishes there are four
too few equations and the system is underdetermined. Thus, we expect four
arbitrary functions in the solution. Such a solution is given by
\begin{equation}
\nabla_{\alpha}\left[\sqrt{g}\left(g^{\mu\nu}+\kappa bT^{\mu\nu}\right)\right]=
\sqrt{g}\left[\delta^{\mu}_{\alpha}V^{\nu}+\delta^{\nu}_{\alpha}
V^{\mu}\right]
\label{matter}
\end{equation}
where $V^{\mu}$ is an arbitrary vector.
  
Now consider vacuum solutions with $T^{\mu\nu}=0$. Contracting equation (\ref{matter})
with $g_{\mu\nu}$ and using
\begin{equation}
\sqrt{g}g_{\mu\nu}\nabla_{\alpha}g^{\mu\nu}=\sqrt{g}\left[g_{\mu\nu}\partial
_{\alpha}g^{\mu\nu}+2\Gamma^{\mu}_{\alpha\mu}\right]=-2\nabla_{\alpha}\sqrt{g}
\end{equation}
gives
\begin{equation}
\nabla_{\alpha}\sqrt{g}=\sqrt{g}V_{\alpha}\; .
\end{equation}
Thus, we have
\begin{equation}
\nabla_{\alpha}g_{\mu\nu}=g_{\mu\nu}V_{\alpha}-g_{\mu\alpha}V_{\nu}-
g_{\nu\alpha}V_{\mu}\; .
\end{equation}
The same procedure that is used in general relativity to find the connection
gives
\begin{equation}
\Gamma^{\alpha}_{\mu\nu}=\left\{
\begin{array}{ll}
\alpha\\
\mu\nu\\
\end{array}
\right\}+\frac{1}{2}\left[3g_{\mu\nu}V^{\alpha}-\delta^{\alpha}_{\mu}V_{\nu}-
\delta^{\alpha}_{\nu}V_{\mu}\right]\; ,
\label{connection}
\end{equation}
where the first term on the right hand side is the Christoffel symbol.
Now go back to equation (\ref{anti}) and substitute in equations 
(\ref{eq1}) and (\ref{matter}) to get (recall that $T_{\mu\nu}=0$)
\begin{equation}
\nabla_{\alpha}\left[\sqrt{P}\left(P^{-1}\right)^{\stackrel{\alpha\mu}{\vee}}\right]
=-3\sqrt{g}V^{\mu}\; .
\label{eq18}
\end{equation}
Now consider the ``Einstein limit'' where $|bR_{\mu\nu}|<<|g_{\mu\nu}|$.
To second order in $b$
\begin{equation}
\left(P^{-1}\right)^{\mu\nu}=g^{\mu\nu}-bR^{\mu\nu}+b^2R^{\mu}_{\;\;\alpha}R^{\alpha\nu}
\end{equation}
and
\begin{equation}
\sqrt{P}=\sqrt{g}\left[1+\frac{1}{2}bR+\frac{1}{8}b^2R^2-\frac{1}{4}b^2R_{\alpha\beta}
R^{\beta\alpha}\right]\; .
\end{equation}
Equation (\ref{eq1}) becomes
\begin{equation}
G{\underline{\mu\nu}}=-\kappa\left(\overline{S}_{\mu\nu}+\hat{S}_{\mu\nu}\right)
\label{Ein1}
\end{equation}
where
\begin{equation}
\overline{S}_{\mu\nu}=\frac{b}{2\kappa}\left[RR_{\underline{\mu\nu}}-\frac{1}{4}R^2g_{\mu\nu}
-2g^{\alpha\beta}R_{\underline{\mu\alpha}}R_{\underline{\nu\beta}}+\frac{1}{2}
g_{\mu\nu}R_{\underline{\alpha\beta}}R^{\underline{\alpha\beta}}\right]\;,
\end{equation}
\begin{equation}
\hat{S}_{\mu\nu}=\frac{b}{\kappa}\left[R_{\stackrel{\mu\alpha}{\vee}}R_{\stackrel{
\nu\beta}{\vee}}g^{\alpha\beta}-\frac{1}{4}g_{\mu\nu}R_{\stackrel{\alpha\beta}{\vee}}
R^{\stackrel{\alpha\beta}{\vee}}\right]\; ,
\end{equation}
$R=g^{\mu\nu}R_{\underline{\mu\nu}}$ 
and $G_{\mu\nu}$ is the Einstein tensor with respect to the connection $\Gamma^{\alpha}
_{\mu\nu}$. Equation (\ref{eq18}) becomes
\begin{equation}
\nabla_{\alpha}\left\{\sqrt{g}\left[\left(1+\frac{1}{2}bR\right)R^{\stackrel{\alpha\mu}{\vee}}
+bR^{\beta[\alpha}R^{\mu]}_{\beta}\right]\right\}
=\frac{3}{b}\sqrt{g}V^{\mu}
\label{em1}
\end{equation}
where the square brackets around $\alpha$ and $\mu$ denote antisymmetrization
on these indices. The antisymmetric part of the Ricci tensor is proportional
to the curl of the vector field and is given by
\begin{equation}
R_{\stackrel{\mu\nu}{\vee}}=\frac{1}{2}\left[\nabla_{\mu}V_{\nu}-\nabla_{\nu}
V_{\mu}\right]\; .
\label{curl}
\end{equation}
Note that to lowest order in $b$ equations (\ref{em1}) and (\ref{curl}) describe
a massive vector field with mass $\sqrt{6/b}$. This implies that we must
take $b$ to be positive.
Equations (\ref{Ein1}) to (\ref{curl}) plus (\ref{connection})
are the field equations of the theory. 
To compare to the Einstein field equations we need to write these equations
in terms of the Einstein tensor $\tilde{G}_{\mu\nu}$, which is defined
in terms of the Christoffel symbol. The relationship between the Ricci tensors
is given by
\begin{equation}
R_{\mu\nu}=\tilde{R}_{\mu\nu}-\tilde{\nabla}_{\alpha}H^{\alpha}_{\mu\nu}+\tilde{\nabla}
_{\nu}H^{\alpha}_{\alpha\mu}-H^{\alpha}_{\alpha\beta}H^{\beta}_{\mu\nu}+H^{\alpha}
_{\mu\beta}H^{\beta}_{\alpha\nu}
\end{equation}
where $H^{\alpha}_{\mu\nu}$ is the tensor field
\begin{equation}
H^{\alpha}_{\mu\nu}=\Gamma^{\alpha}_{\mu\nu}-\left\{
\begin{array}{ll}
\alpha\\
\mu\nu\\
\end{array}
\right\}
\end{equation}
and $\tilde{\nabla}$ is the metric compatible covariant derivative.
A simple calculation shows that
\begin{equation}
G_{\underline{\mu\nu}}=\tilde{G}_{\mu\nu}+\frac{3}{2}\left[V_{\mu}V_{\nu}+\frac{1}{2}
g_{\mu\nu}\left(2\tilde{\nabla}_{\alpha}V^{\alpha}-V_{\alpha}V^{\alpha}\right)
\right]\; .
\label{G}
\end{equation}
It is easy to simplify the right hand side by showing that $\tilde{\nabla}_{\alpha}
V^{\alpha}=0$. To begin take the divergence of equation (\ref{em1}) and 
use (\ref{matter}) to get
\begin{equation}
\nabla_{\mu}\nabla_{\alpha}R^{\stackrel{\alpha\mu}{\vee}}+\nabla_{\mu}\left(V_{\alpha}
R^{\stackrel{\alpha\mu}{\vee}}\right)+\nabla_{\mu}U^{\mu}=\frac{3}{b}\nabla_{\mu}V^{\mu}
\; ,
\label{int}
\end{equation} 
where $U^{\mu}$ contains the terms in (\ref{em1}) proportional to $b$. A
short calculation shows that
\begin{equation}
\nabla_{\mu}\nabla_{\alpha}R^{\stackrel{\alpha\mu}{\vee}}=R_{\stackrel{\mu
\alpha}{\vee}}R^{\stackrel{\mu\alpha}{\vee}}
\label{div1}
\end{equation}
and that $\nabla_{\mu}U^{\mu}$ is third order in the curvature so that 
it can be neglected. Substituting (\ref{div1}) and (\ref{em1}) into
(\ref{int}) and neglecting third order terms in the curvature gives
\begin{equation}
\nabla_{\mu}V^{\mu}=-V^{\alpha}V_{\alpha}\; .
\end{equation}
Using (\ref{connection}) it is easy to show that
\begin{equation}
\tilde{\nabla}_{\mu}V^{\mu}=0\; .
\end{equation}
Now define the vector potential $A_{\mu}$ 
by
\begin{equation}
A_{\mu}=\frac{1}{2\alpha}V_{\mu}\; .
\end{equation}
The antisymmetric part of the Ricci tensor is then given by
\begin{equation}
R_{\stackrel{\mu\nu}{\vee}}=\alpha F_{\mu\nu}
\end{equation}
where $F_{\mu\nu}=\tilde{\nabla}_{\mu}A_{\nu}-\tilde{\nabla}_{\nu}A_{\mu}$.
Note that the $\nabla$ operator in (\ref{curl}) can be replaced by $\tilde
{\nabla}$ or by $\partial$. 
   
To summarize field equations are given by
\begin{equation}
\tilde{G}_{\mu\nu}=-\kappa \left[T^{R}_{\mu\nu}+T^{A}_{\mu\nu}\right]
\label{ein2}
\end{equation}
and by
\begin{equation}
\tilde{\nabla}_{\alpha}F^{\alpha\mu}-\frac{6}{b}A^{\mu}=S^{\mu}
\label{em2}
\end{equation}
where
\begin{equation}
T^{R}_{\mu\nu}=\frac{b}{2\kappa}\left[RR_{\underline{\mu\nu}}-\frac{1}{4}R^2g_{\mu\nu}
-2g^{\alpha\beta}R_{\underline{\mu\alpha}}R_{\underline{\nu\beta}}+\frac{1}{2}
g_{\mu\nu}R_{\underline{\alpha\beta}}R^{\underline{\alpha\beta}}\right]\;,
\label{TR}
\end{equation}
\begin{equation}
T^{A}_{\mu\nu}=\frac{\alpha^2b}{\kappa}\left[F_{\mu\alpha}
F^{\;\;\alpha}_{\nu}-\frac{1}{4}g_{\mu\nu}F^{\alpha\beta}F_{\alpha\beta}
+\frac{6}{b}\left(A_{\mu}A_{\nu}-\frac{1}{2}g_{\mu\nu}A^{\alpha}A_{\alpha}
\right)\right]\; ,
\label{TF}
\end{equation}
and $S^{\mu}$ contains the terms in (\ref{em1}) that are proportional to
$b$. Note that equation (\ref{em2})
describes a vector field with mass $\sqrt{6/b}$ and $T^{A}_{\mu\nu}$ is its
corresponding energy momentum tensor.
  
As discussed below equation (\ref{Ricci}) the vacuum field equations with
$F_{\mu\nu}=0$ are the Einstein equations for all values of $b$. Thus, 
vacuum tests of general relativity do not constrain the value of $b$. We
will see below however that the value of $b$ is severely constrained by 
taking general relativity to be valid in the interior of neutron stars.
  
Now consider the field equations in the presence of matter and sources for
the vector field. To the Lagrangian given in (\ref{lag}) we need to add an
interaction term that couples the vector field to its sources. The simplest
coupling is given by
\begin{equation}
L_c=\frac{\alpha b}{2\kappa}\sqrt{g}\;\Gamma^{\alpha}_{\alpha\mu}
J^{\mu}\; 
\label{L2}
\end{equation}
where $J^{\mu}$ is the conserved current associated with the source.
At first sight it may appear that there is a problem with this choice as
$\Gamma^{\alpha}_{\alpha\mu}$ is not a tensor. An analogous situation occurs
in electrodynamics where the interaction Lagrangian $\sqrt{g}A_{\mu}J^{\mu}$
appears not to be gauge invariant. However, if $\tilde{\nabla}_{\mu}J^{\mu}
=0$ the Lagrangian only changes by a total derivative under a gauge
transformation. Now, under a coordinate transformation
\begin{equation}
\bar{\Gamma}^{\alpha}_{\alpha\mu}=\frac{\partial x^{\nu}}
{\partial \bar{x}^{\mu}}\Gamma^{\alpha}_{\alpha\nu}-\frac{\partial}{\partial
\bar{x}^{\mu}}\ln\left|\frac{\partial x}{\partial\bar{x}}\right|
\end{equation}
where $|\partial x/\partial\bar{x}|$ is the Jacobian of the transformation.
This is analogous to a gauge transformation and it is easy to show
that the Lagrangian $L_c$ only changes by a total derivative
if $\tilde{\nabla}_{\mu}J^{\mu}=0$. Another way of looking at it
is to use (\ref{connection}) to write 
\begin{equation}
\sqrt{g}\;\Gamma^{\alpha}_{\alpha\mu}J^{\mu}=\sqrt{g}\left[\partial_{\mu}
\ln\sqrt{g}-V_{\mu}\right]J^{\mu}\; .
\end{equation}
The first term can be rewritten as a total divergence and can therefore
be neglected. The second term is a scalar density and so the Lagrangian
$L_c$ has the correct transformation properties. A simple example of a
conserved current is a point source with charge $e$. The current
is given by
\begin{equation}
J^{\mu}(x^{\alpha})=\frac{e}{\sqrt{g}}\int U^{\mu}(\tau)\delta(x^{\alpha}
-x^{\alpha}(\tau))d\tau
\end{equation}
where $U^{\mu}$ is the four velocity of the particle and $\tau$ is the
proper length along its world line. Of course the fields that produce
$J^{\mu}$ must appear in $L_M$.
One interesting
property of this choice of Lagrangian is that the current $J^{\mu}$
does not enter into the equation defining the connection, so 
that (\ref{connection}) is still valid. In the Einstein limit where
$b|R_{\mu\nu}|<<|g_{\mu\nu}|$ and $\kappa b|T^{\mu\nu}|<<|g^{\mu\nu}|$
the field equations are given by
\begin{equation}
\tilde{G}_{\mu\nu}=-\kappa \left[T_{\mu\nu}+T^{R}_{\mu\nu}+T^{A}_{\mu\nu}
+T^{\Gamma}_{\mu\nu}\right]
\label{ein2}
\end{equation}
and
\begin{equation}
\tilde{\nabla}_{\alpha}F^{\alpha\mu}-\frac{6}{b}A^{\mu}=J^{\mu}+S^{\mu}
\label{em20}
\end{equation}
where $T^{R}_{\mu\nu}$ is given by (\ref{TR}), $T^{A}_{\mu\nu}$ is
given by (\ref{TF}), $S^{\mu}$ contains the terms proportional to
$b$ in (\ref{em1}) and $T^{\Gamma}_{\mu\nu}$ contains the additional
terms that enter through the connection from the $\kappa bT^{\mu\nu}$ 
terms in (\ref{last}).
  
For the Einstein equations to be approximately valid it is necessary that
$\kappa b|T^{\mu\nu}|<<|g^{\mu\nu}|$. If $g_{\mu\nu}\sim\eta_{\mu\nu}$ and
if $T^{\mu\nu}$ describes matter with a density $\rho$ then the constraint
can be written as
\begin{equation}
b<<\frac{1}{\kappa\rho}\; .
\end{equation}
For a neutron star $\rho\simeq 10^{18}$ kg/m$^3$ and the constraint becomes
\begin{equation}
b<<10^{8}\;  m^2\; .
\label{constraint}
\end{equation}
This corresponds to the mass constraint
\begin{equation}
m>>10^{-47}\; kg\; .
\end{equation}
Since the mass of the photon is constrained to be less than $10^{-52}$ kg
 \cite{Wi1,Da1,Gr1} this vector field cannot be the electromagnetic field.
    
The field equations (\ref{ein2}) can be derived from the Lagrangian
\begin{equation}
L=-\frac{1}{2\kappa}\sqrt{g}\left[ R+\frac{1}{4}bR^2-\frac{1}{2}b
R_{\alpha\beta}R^{\beta\alpha}\right]+L_c+L_M
\end{equation}
which is the Born-Infeld-Einstein Lagrangian (\ref{lag}) expanded to order
$b$. Equation (\ref{em2}) also follows, but without the $S^{\mu}$ term
which is of order $ b^2$ in the expansion of the Lagrangian. Thus,
this term can be consistently neglected to this order.
\section*{Field Equations II}
On a manifold with a general connection there are two nonzero contractions
of the Riemann tensor: $R_{\mu\nu}$ given in (\ref{Ricci1}) and
\begin{equation}
S_{\mu\nu}=R^{\alpha}_{\;\;\alpha\mu\nu}=\partial_{\mu}\Gamma^{\alpha}_
{\alpha\nu}-\partial_{\nu}\Gamma^{\alpha}_{\alpha\mu}\; .
\label{S}
\end{equation}
The simplest Born-Infeld Lagrangian that includes both $R_{\mu\nu}$ and
$S_{\mu\nu}$ is given by replacing $bR_{\mu\nu}$ in (\ref{lag}) by the 
linear combination $bR_{\mu\nu}+dS_{\mu\nu}$.
Since $S_{\mu\nu}=-2R_{\stackrel{\mu\nu}{\vee}}$ this Lagrangian can be 
written as
\begin{equation}
L=-\frac{1}{\kappa b}\left\{\sqrt{-det(g_{\mu\nu}+bR_{\underline{\mu\nu}}+aR_{\stackrel
{\mu\nu}{\vee}})}-\sqrt{-det(g_{\mu\nu})}\right\}+L_M\; .
\label{lag2}
\end{equation}
If $a=b$ we obtain the theory discussed in the previous section.
Varying the action with respect to $g_{\mu\nu}$
gives
\begin{equation}
\sqrt{P}\left(P^{-1}\right)^{\underline{\mu\nu}}-\sqrt{g}g^{\mu\nu}
=\sqrt{g}\kappa bT^{\mu\nu}
\label{eq1a}
\end{equation}
where $P_{\mu\nu}=g_{\mu\nu}+bR_{\underline{\mu\nu}}+aR_{\stackrel{\mu\nu}{\vee}}$.
Varying with respect to 
$\Gamma^{\alpha}_{\mu\nu}$ gives
\begin{equation}
\nabla_{\alpha}\left[\sqrt{P}\left(P^{-1}\right)^{\underline{\mu\nu}}\right]
-\frac{1}{2}\nabla_{\beta}\left\{\sqrt{P}\left[
\left(P^{-1}\right)^{\underline{\mu\beta}}+\frac{a}{b}\left(P^{-1}\right)^{\stackrel{\mu\beta}{\vee}}
\right]\delta^{\nu}_{\alpha}+\sqrt{P}\left[
\left(P^{-1}\right)^{\underline{\nu\beta}}+\frac{a}{b}\left(P^{-1}\right)^{\stackrel{\nu\beta}{\vee}}
\right]\delta^{\mu}_{\alpha}
\right\}=0
\label{eq2c}
\end{equation}
and contracting over $\alpha$ and $\nu$ gives
\begin{equation}
\nabla_{\alpha}\left[\sqrt{P}\left(P^{-1}\right)^{\stackrel{\alpha\mu}
{\vee}}\right]=-\frac{3b}{5a}\nabla_{\alpha}\left[\sqrt{P}\left(P^{-1}
\right)^{\underline{\alpha\mu}}\right]\;
\label{anti2}
\end{equation}
As before, the field equations reduce to the Einstein field equations 
in a vacuum if $R_{\stackrel{\mu\nu}{\vee}}=0$. Substituting equations 
(\ref{eq1a}) and (\ref{anti2}) into (\ref{eq2c}) gives (\ref{last}).
Once again set
\begin{equation}
\nabla_{\alpha}\left[\sqrt{g}\left(g^{\mu\nu}+\kappa bT^{\mu\nu}\right)\right]=
\sqrt{g}\left[\delta^{\mu}_{\alpha}V^{\nu}+\delta^{\nu}_{\alpha}
V^{\mu}\right]
\label{matter1}
\end{equation}
where $V^{\mu}$ is an arbitrary vector and the connection is still 
given by (\ref{connection}). Now substitute equations (\ref{eq1a}) 
and (\ref{matter1}) into (\ref{anti2}) to get
\begin{equation}
\nabla_{\alpha}\left[\sqrt{P}\left(P^{-1}\right)^{\stackrel{\alpha\mu}{\vee}}\right]
=-\frac{3b}{a}\sqrt{g}V^{\mu}\; .
\label{eq18a}
\end{equation}
Now consider the ``Einstein limit'' where $b|R_{\mu\nu}|<<|g_{\mu\nu}|$.
To second order in $b$ and $a$
\begin{equation}
\left(P^{-1}\right)^{\underline{\mu\nu}}=g^{\mu\nu}-bR^{\underline{\mu\nu}}+
b^2g_{\alpha\beta}R^{\underline{\mu\alpha}}R^{\underline{\beta\nu}}
+a^2g_{\alpha\beta}R^{\stackrel{\mu\alpha}{\vee}}R^{\stackrel{\beta\nu}{\vee}}
\; ,
\end{equation}
\begin{equation}
\left(P^{-1}\right)^{\stackrel{\mu\nu}{\vee}}=-a\left[R^{\stackrel{\mu\nu}{\vee}}
-b\left(R^{\underline{\mu\alpha}}R^{\stackrel{\beta\nu}{\vee}}+R^{\stackrel{\mu\alpha}
{\vee}}R^{\underline{\beta\nu}}\right)\right]
\end{equation}
and
\begin{equation}
\sqrt{P}=\sqrt{g}\left[1+\frac{1}{2}bR+\frac{1}{8}b^2R^2-\frac{1}{4}b^2
R_{\underline{\alpha\beta}}R^{\underline{\alpha\beta}}+\frac{1}{4}a^2
R_{\stackrel{\beta\alpha}{\vee}}R^{\stackrel{\alpha\beta}{\vee}}
\right]\; ,
\end{equation}
where $R=g^{\mu\nu}R_{\underline{\mu\nu}}$.
Sources can be coupled to the vector field by using
the interaction Lagrangian $L_c$ given in (\ref{L2})
with $b$ replaced by $a$.
Using equation (\ref{G}) and the above gives
\begin{equation}
\tilde{G}_{\mu\nu}=-\kappa \left[T_{\mu\nu}+T^{R}_{\mu\nu}+T^{A}_{\mu\nu}
+T^{\Gamma}_{\mu\nu}\right]
\label{ein2a}
\end{equation}
and
\begin{equation}
\tilde{\nabla}_{\alpha}F^{\alpha\mu}-\frac{6b}{a^2}A^{\mu}=J^{\mu}+S^{\mu}
\label{em2a}
\end{equation}
where
\begin{equation}
T^{R}_{\mu\nu}=\frac{b}{2\kappa}\left[RR_{\underline{\mu\nu}}-\frac{1}{4}R^2g_{\mu\nu}
-2g^{\alpha\beta}R_{\underline{\mu\alpha}}R_{\underline{\nu\beta}}+\frac{1}{2}
g_{\mu\nu}R_{\underline{\alpha\beta}}R^{\underline{\alpha\beta}}\right]\;,
\label{TRa}
\end{equation}
\begin{equation}
T^{A}_{\mu\nu}=\frac{\alpha^2c^2}{b\kappa}\left[F_{\mu\alpha}
F^{\;\;\alpha}_{\nu}-\frac{1}{4}g_{\mu\nu}F^{\alpha\beta}F_{\alpha\beta}
+\frac{6b}{a^2}\left(A_{\mu}A_{\nu}-\frac{1}{2}g_{\mu\nu}A^{\alpha}A_{\alpha}
\right)\right]
\; ,
\label{TFa}
\end{equation}
and $S^{\mu}$ contains the terms in (\ref{em1}) that are proportional to
$b$. It is easy to show that the constraint $\tilde{\nabla}_{\mu}A^{\mu}=0$
is still satisfied. Note that equation (\ref{em2a})
describes a vector field with mass $\sqrt{6b/a^2}$ and $T^{A}_{\mu\nu}$ 
is its corresponding energy momentum tensor.
  
The value of $b$ is constrained by
\begin{equation}
b<< 10^8\;\; m^2
\end{equation}
if we require that general relativity holds to a good approximation inside
neutron stars. The  mass of the photon is constrained to be less than 
$10^{-52}$ kg \cite{Wi1,Da1,Gr1} and this gives the constraint
\begin{equation}
a>10^{10}\sqrt{b}\; .
\end{equation}
Thus, the coupling constant associated with the antisymmetric part of the
Ricci tensor must be much larger than the coupling constant associated
with the symmetric part.
  
The field equations without the $S^{\mu}$ term can be derived from the
Lagrangian
\begin{equation}
L=-\frac{1}{2\kappa}\sqrt{g}\left[R+\frac{1}{4}bR^2-\frac{1}{2}bR_{\underline{\alpha\beta}}
R^{\underline{\alpha\beta}}+\frac{c^2}{2b}R_{\stackrel{\alpha\beta}{\vee}}
R^{\stackrel{\alpha\beta}{\vee}}\right]+L_c+L_M\;
\label{explag}
\end{equation}
which is the expansion of (\ref{lag2}) to first order in $b$ and $c^2/b$.
\section*{Curvature squared Lagrangian}
In the previous two sections it was shown that Born-Infeld-Einstein
Lagrangians give gravity coupled to a massive vector field to first
order in the parameters $b$ and $c^2/b$. In this section a 
Lagrangian which exactly produces the Einstein field equations
coupled to a massive vector field will be given.
  
The Lagrangian for the theory in the absence of sources is
\begin{equation}
L=-\frac{1}{2\kappa}\sqrt{g}\left( R+\frac{1}{2}bR_{\stackrel{\alpha\beta}{\vee}}
R^{\stackrel{\alpha\beta}{\vee}}\right)\; .
\end{equation}
This is (\ref{explag}) with the quadratic terms in $R_{\underline{\mu\nu}}$
dropped. 
  
Varying with respect to $g_{\mu\nu}$ gives
\begin{equation}
G_{\mu\nu}=-\alpha^2b\left[ F_{\mu\alpha}F_{\nu}^{\;\;\alpha}-\frac{1}{4}
g_{\mu\nu}F_{\stackrel{\alpha\beta}{\vee}}F^{\stackrel{\alpha\beta}{\vee}}
\right]\; .
\label{einf}
\end{equation}
Varying with respect to $\Gamma^{\alpha}_{\mu\nu}$ one finds that the
connection is again given by (\ref{connection}) and the field equation for
$F^{\mu\nu}$ is given by
\begin{equation}
\tilde{\nabla}_{\alpha}F^{\alpha\mu}=\frac{6}{b}A^{\mu}
\end{equation}
Converting $G_{\mu\nu}$ to $\tilde{G}_{\mu\nu}$ in (\ref{einf}) gives the
Einstein field equations with the energy momentum tensor (\ref{TF}). Thus,
the field equations describe a massive vector field coupled to gravity.
\section*{Conclusion}
The field equations for the Born-Infeld-Einstein action were derived 
using a Palatini variation. The vacuum field equations with $R_
{\stackrel{\mu\nu}{\vee}}=0$ were shown to be the Einstein vacuum
equations, independent of the value of $b$. This extends the universality
property of the vacuum Einstein field equations discovered by Ferraris
et al. to the Born-Infeld-Einstein action. For sufficiently small curvatures
the field equations describe a massive vector field with $R_{\stackrel{\mu\nu}
{\vee}}$ being proportional to the field tensor and the Einstein 
equations plus small corrections. The simplest version of the theory uses only the Ricci tensor
$R_{\mu\nu}$. By requiring that the Einstein field equations hold to a good
approximation in neutron stars
it was shown that the mass of the vector field exceeds the
limit on the photon mass. Thus, in this case the vector field would necessarily
describe a new field. 
  
On a manifold with a general connection
there are two nonzero contractions of the Riemann tensor: $R_{\mu\nu}$ and
$S_{\mu\nu}$.
If both $R_{\mu\nu}$ and $S_{\mu\nu}$ are used it is possible
to have the Einstein equations hold in neutron stars and for the mass
of the field to satisfy the constraints on the photon mass. For this to work
the coupling constant associated with the antisymmetric part of the Ricci
tensor must be much larger than the coupling constant associated with the
symmetric part. 

\end{document}